# Processing Dynamics of 3D-Printed Carbon Nanotubes-Epoxy Composites

Ali Khater[†,1], Sohini Bhattacharyya[†,1], M. A. S. R. Saadi[1], Morgan Barnes[1], Minghe Lou[2], Vijay Harikrishnan[1], Seyed Mohammad Sajadi[1], Peter J. Boul[3], Chandra Sekhar Tiwary[4], Hanyu Zhu[1], Muhammad M. Rahman[*,1], Pulickel M. Ajayan[*,1]

[1]Department of Materials Science and NanoEngineering, Rice University, Texas, USA
[2]Department of Electrical and Computer Engineering, Rice University, Texas, USA
[3]Aramco Americas, Texas, USA
[4]Metallurgical and Materials Engineering, Indian Institute of Technology Kharagpur, India

[*]Corresponding author
[†]These authors contributed equally.

**Abstract**

**Carbon Nanotubes (CNTs)-polymer composites are promising candidates for a myriad of applications. Ad-hoc CNTs-polymer composite fabrication techniques inherently pose roadblock to optimized processing resulting in microstructural defects i.e., void formation, poor interfacial adhesion, wettability, and agglomeration of CNTs inside the polymer matrix. Although improvement in the microstructures can be achieved via additional processing steps such as-mechanical methods and/or chemical functionalization, the resulting composites are somewhat limited in structural and functional performances. Here, we demonstrate that 3D printing technique like- direct ink writing offers improved processing of CNTs-polymer composites. The shear-induced flow of an engineered nanocomposite ink through the micronozzle offers some benefits including reducing the number of voids within the epoxy, improving CNTs dispersion and adhesion with epoxy, and partially aligns the CNTs. Such microstructural changes result in superior mechanical performance and heat transfer in the composites compared to their mold-casted counterparts. This work demonstrates the advantages of 3D printing over traditional fabrication methods, beyond the ability to rapidly fabricate complex architectures, to achieve improved processing dynamics for fabricating CNT-polymer nanocomposites with better structural and functional properties.**

## 1. Introduction

Carbon nanotubes (CNTs) demonstrate remarkable mechanical, thermal, electronic, and optical properties resulting in their widespread use across many disciplines.[1–4] Particularly, CNTs are promising fillers as mechanical reinforcement in polymer composites because of their high modulus, strength, specific surface area, and aspect ratio.[5,6] However, CNTs composites for structural applications have been limited due to poor interfacial adhesion, wettability, and agglomerations of the CNTs filler within the host material.[7–10] These challenges are difficult to overcome since the very same properties that make CNTs promising lead to these inherent flaws. For instance, while the high specific surface area of CNTs provides desirable interface for stress transfer in polymer composites, it introduces strong attractive forces among CNTs causing excessive agglomeration and produces unwanted stress concentrations for failure.[11,12] This makes it difficult to develop well dispersed CNTs composites due to their strong tendency to re-agglomerate. Various mechanical methods to disperse nanotubes in polymer, such as ultra-sonication, ball-milling, and high shear mixing have been adopted.[13–15] Also, chemical functionalization of the CNTs surface improved interfacial interaction between CNTs and matrix, and the dispersion of CNTs into the matrix.[16–18] However, these processes have limitations and can present several challenges, e.g., void formation during mold casting, high viscosity, limited processability, and eventually low mechanical strength.

Three-dimensional (3D) printing is a rapidly growing manufacturing technology due to its versatility and ability to produce macroscopically complex structures with controlled microstructures and properties.[19–21] Particularly, direct ink writing (DIW) can print composites with intricate architectures using high viscosity materials that eliminates the need of any mold or template.[22–25] The primary basis of DIW technique is the development of viscoelastic inks, with appropriate shear-thinning properties, to be extruded by a micronozzle at suitably low pressures, while possessing sufficiently high storage modulus and yield strength to hold the printed construct.[26] It allows superior control over the composition, shape, and geometry of complex polymer structures, giving it an edge over traditional fabrication methods. Additionally, this extrusion-based technique is favored because shear-alignment during printing can be leveraged to align fillers during extrusion resulting in anisotropic printed architectures.[27–29] Since shear-induced flow is intrinsically coupled to the DIW process, it is necessary to understand the effect of shear stresses on void formation, particle dispersion, and alignment of anisotropic particles during printing. Understanding the dynamics of such issues during the flow of non-Newtonian fluids is critically important to design proper ink materials and 3D printed structures with better physical properties such as- mechanical and thermal.

Here, we develop CNT-reinforced epoxy nanocomposite ink and 3D-print architected structures and compare them with structures developed by conventional mold-casting approach to gain insight into the effect of printing through micro-nozzles on the structures and properties of the printed structures. The tapered micro-nozzle used for extrusion imparts a high shear stress on the ink, reducing the voids in the structure that are responsible for crack initiation and thus provides greater mechanical strength to the structure. Also, the 3D printing facilitates better dispersion, alignment, and wetting of the CNT fillers in the composite structures (**Figure 1(a)**) along with the removal of structural supports, reduction of material waste and further processing to reduce voids as required during mold casting. Using optical and electron microscopic techniques together with Raman spectroscopy and theoretical simulations, we provide evidence of the improved processing dynamics e.g., void reduction, better dispersion, partial alignment, and wetting of CNTs during direct writing of CNTs-epoxy composites through a tapered nozzle. While processing of nanocomposites at higher CNT loading following a traditional approach is difficult due to high viscosity build-up, our approach in this work demonstrate that the printing technique eliminates such processing challenges. Thus, better wetted, and dispersed CNTs in epoxy resin results in superior mechanical and thermal performance in 3D-printed structures as compared to their traditional counterparts.

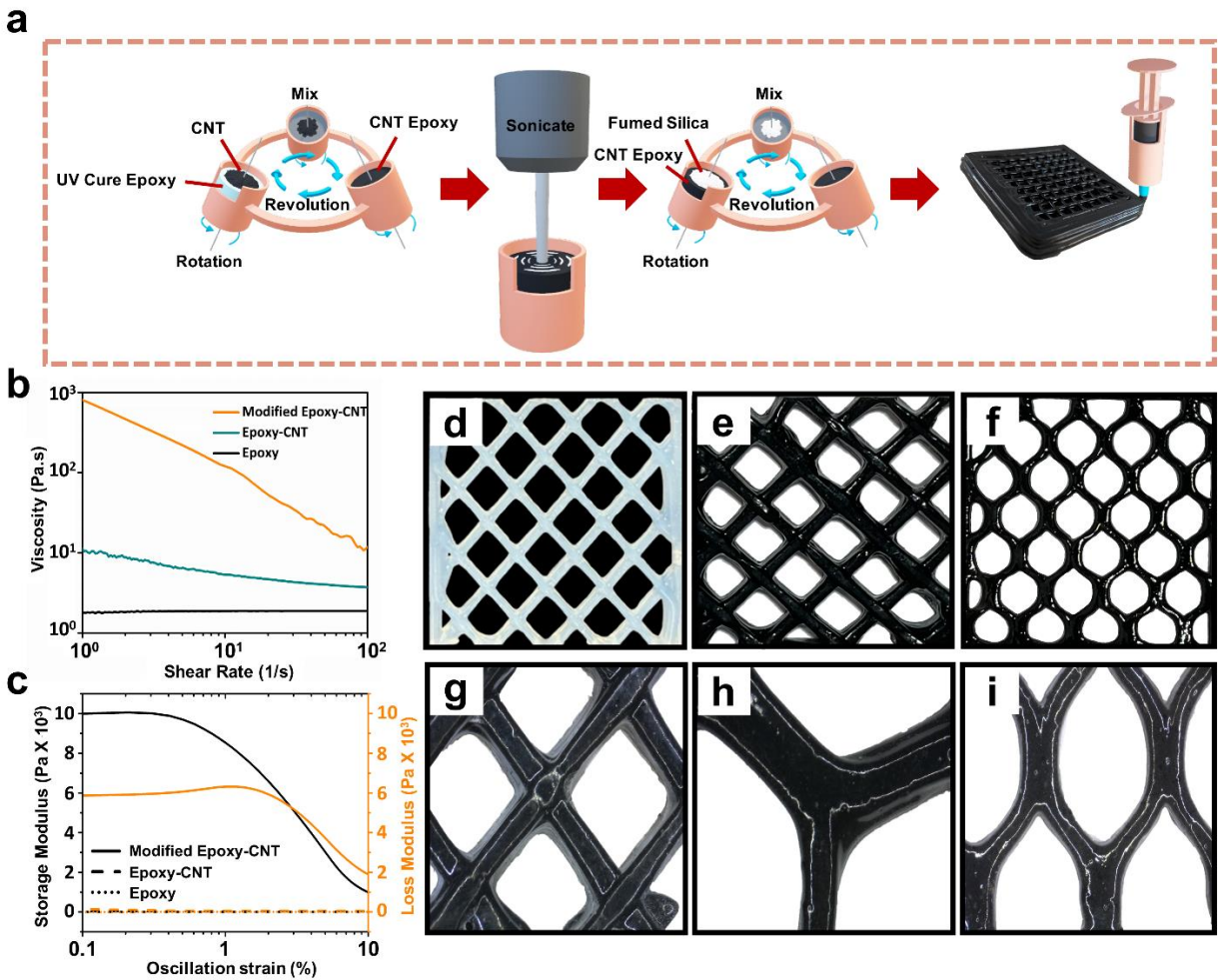

**Figure 1.** **(a)** Schematic diagram depicting the various steps of the fabrication process of CNT-epoxy composite using direct ink writing technique. **(b)** Apparent viscosity as a function of shear rate of 3D-printable (modified) CNT-Epoxy, CNT-Epoxy, and Epoxy. **(c)** Storage and loss modulus of 3D-printable CNT-Epoxy, CNT-Epoxy, and Epoxy as a function of oscillation strain. **(d)** Photograph of 3D-printed rectilinear structure composed of epoxy polymer. **(e-f)** Photographs of 3D-printed rectilinear and hexagonal honeycomb structure composed of CNT-epoxy nanocomposites. **(g-i)** Optical microscopy images of nodes of 3D-printed CNT-epoxy nanocomposites.

## 2. Results and discussions

Direct ink writing requires the rheological properties of the CNT-polymer ink to be properly tailored to allow for extrusion at low pressures while maintaining its shape after printing. We use an epoxy resin as the host materials because epoxies are widely regarded for their high stiffness, strength, chemical resistance, low cost, and versatile polymerization routes.[30,31] Pristine UV-cure epoxy resin behaves as a Newtonian fluid and is thus not suitable for DIW because it lacks the necessary shear-thinning behavior. After dispersing different concentrations of CNTs (0, 0.1, 0.3 and 1 wt.%) in the epoxy matrix (**Figure 1a**), the resin demonstrates a considerable increase in both the storage (G′) and loss modulus (G″) and shows slight shear-thinning behavior (**Figure 1b**). However, the CNTs-epoxy ink cannot be printed, at this stage, due to its relatively low viscosity at low shear rates and because the loss modulus is higher than the storage modulus at low shear rates. This indicates that after printing, the material will exhibit more fluid-like behavior rather than solid-like and will not be able to maintain the filamentary shape and/or printed structure. Ideally, there should be a cross-over point between the storage and loss modulus as a function of shear rate or strain so that during extrusion the loss modulus is higher, and the ink acts more fluidic, while after printing at low shear rates/strain, the storage modulus is higher and the ink acts as a solid. In order to resolve these issues, the inks were modified by adding a common thixotropic agent (fumed silica) to obtain rheological behavior suitable for printing. Viscoelastic behavior of the inks under strains akin to DIW reveals that G′ at low strain for fumed silica modified CNT-epoxy ink is one order of magnitude greater than G″, which helps retain the filamentary shape upon extrusion. Moreover, the cross-over point for the modified CNT-epoxy ink occurs at low enough strains to enable (**Figure 1c**) printing using typical extrusion forces of 3D printers. Additionally, the highly shear-thinning behavior of the ink is ideal to yield smooth flow during printing without clogging (**Figure 1b, c**). Following the direct ink writing, the CNTs-epoxy composite is photocured using a broadband 365 nm UV light to permanently solidify the material.

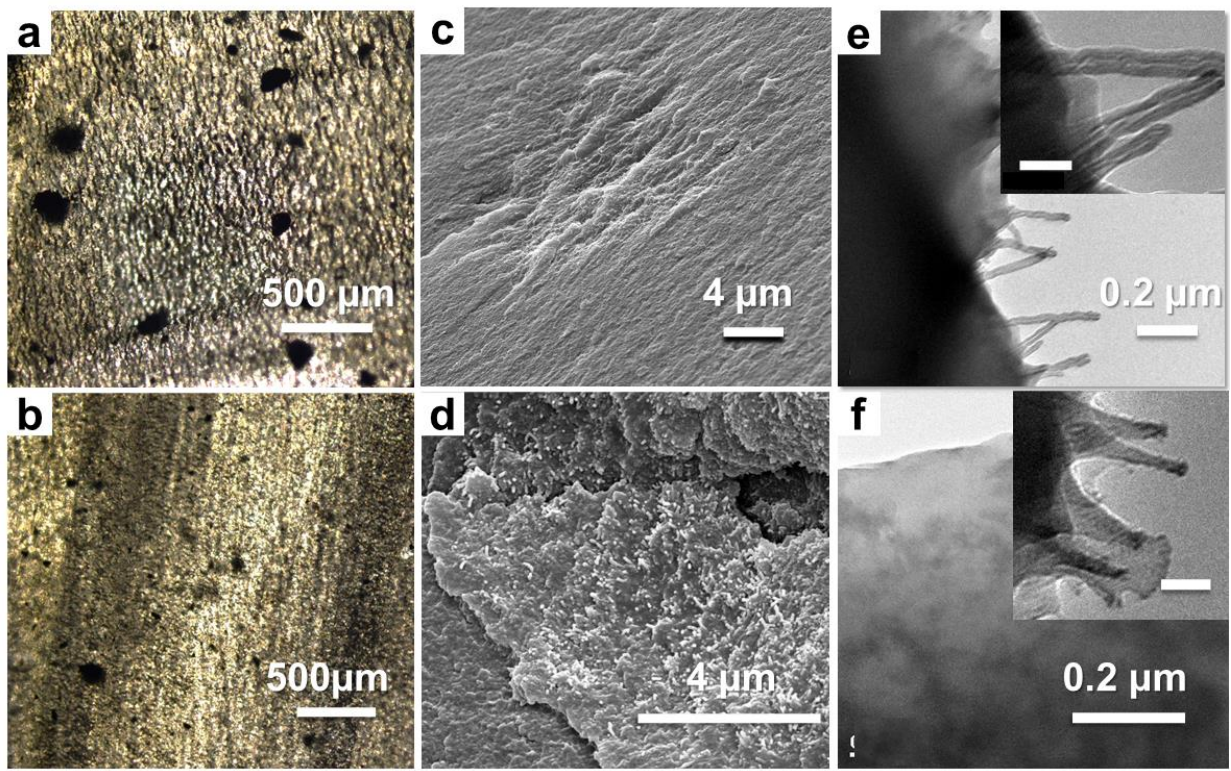

**Figure 2** Optical microscopy images from cross-section of **(a)** mold cast and **(b)** 3D-printed sample. Scanning Electron Microscopy (SEM) images of the fractured surfaces of **(c)** mold cast and **(d)** 3D-printed CNT-epoxy nanocomposite. Transmission Electron Microscopy (TEM) images of the fractured surface of **(e)** mold cast and **(f)** 3D-printed CNT-epoxy nanocomposite.

Next, we evaluated and compared the overall structure and morphology of the 3D-printed and mold-casted CNT-epoxy nanocomposites containing different percentages of CNT (0, 0.1, 0.3 and 1.0 wt.%) in different architectures (**Figure 1d-i**). Typically, epoxies with increasing concentrations of nanofiller materials, such as CNTs, are susceptible to the formation of voids.[6] Strong particle–particle interaction leads to an increase in viscosity in the polymer nanocomposites with increasing CNTs which hinder the removal of entrapped voids from the polymer phase during processing. Using heat, mechanical vibration, or vacuum to remove voids from these highly viscous materials has proven to be ineffective leaving behind voids in the final product which act as stress concentrator for mechanical fracture.[10] However, we observed that 3D printing can reduce the number of voids in the well dispersed CNT-epoxy composite to produce materials with superior performance compared to structures realized via traditional fabrication techniques such as- molding. As expected, mold-cast nanocomposite contains voids as seen through optical microscopy (**Figure 2a**), whereas the printed composite exhibits a surface with minimal voids (**Figure 2b**). The high shear forces exerted during extrusion through the micronozzle is responsible for the significant reduction of voids in the printed structure. Further, the fracture surfaces of mold cast and 3D-printed CNT-epoxy composite were analyzed via Scanning Electron Microscopy (SEM) to understand the differences in their internal microstructures. The fracture surface is relatively smooth in mold cast samples (**Figure 2c**) while that in the printed structures has a greater degree of roughness (**Figure 2d**), indicating continual crack deflection and propagation in different fracture planes.[30] In addition to the microstructure analysis by SEM, we have analyzed the samples using Transmission Electron Microscopy (TEM) to gain insight about the interfaces of CNTs and epoxy. As shown in **Figure 2e,** the mold cast sample reveals large quantity of nanotube pull-out with poor wetting or nanotube-matrix bonding. Conversely, in case of the printed sample, CNTs are covered by epoxy, exhibiting better adhesion, or wetting of the CNT within the epoxy (**Figure 2f**).

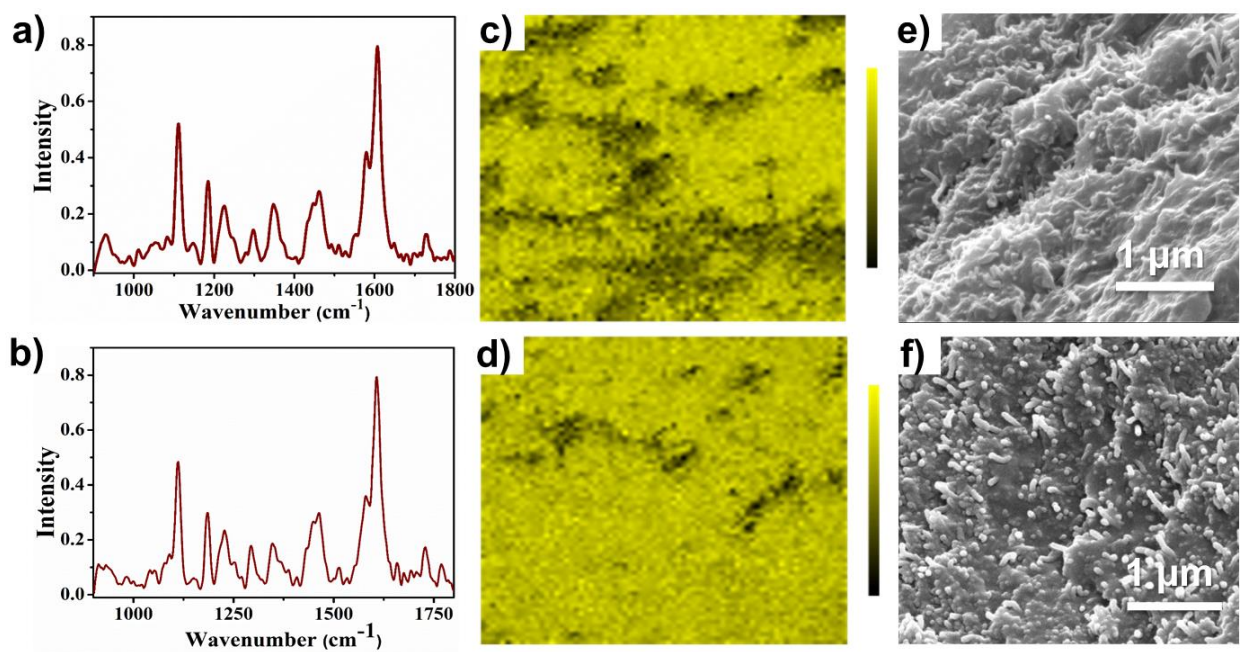

**Figure 3 (a)** Raman spectrum for mold cast CNT-epoxy composite. **(b)** Raman spectrum for 3D printed CNT-epoxy composites. **(c)** Raman mapping of mold cast CNT-epoxy composites. Scale bar indicator using G-peak signal to baseline indicated with yellow for increasing intensity. **(d)** Raman spectrum for 3D-printed CNT-epoxy composites. SEM images of fractured surface of **(e)** mold cast CNT-epoxy composites showing agglomeration of CNT and **(f)** 3D-printed CNT-epoxy composite showing better dispersion and alignment of CNTs.

Raman spectroscopy was used to determine the homogenous dispersion of the CNTs in the epoxy matrix for mold cast and printed samples. Both samples exhibit the typical peaks for CNTs with a D band at 1290 cm$^{-1}$ and a G band at 1600 cm$^{-1}$, in addition to peaks for epoxy and fumed silica (**Figure 3a,b**).[32] However, Raman mapping shows agglomeration of CNTs in the mold cast samples as evident from the optically opaque regions (**Figure 3c**), while the printed sample shows a better dispersion of the CNT (**Figure 3(d)**). This was further corroborated from the SEM image (**Figure 3e**), which shows a random dispersion of CNTs devoid of any alignment in case of the mold cast sample. In contrast, the printed sample showed the presence of dispersed CNTs throughout the epoxy resin (**Figure 3f**). Additionally, the SEM image provides evidence of partial CNTs alignment in the printed sample. This alignment is expected as shear forces and extensional flow field that develops within the micronozzle during printing is known to align anisotropic fillers in a matrix.[27] On the other hand, the mold-casted sample has no external force to induce CNT alignment.

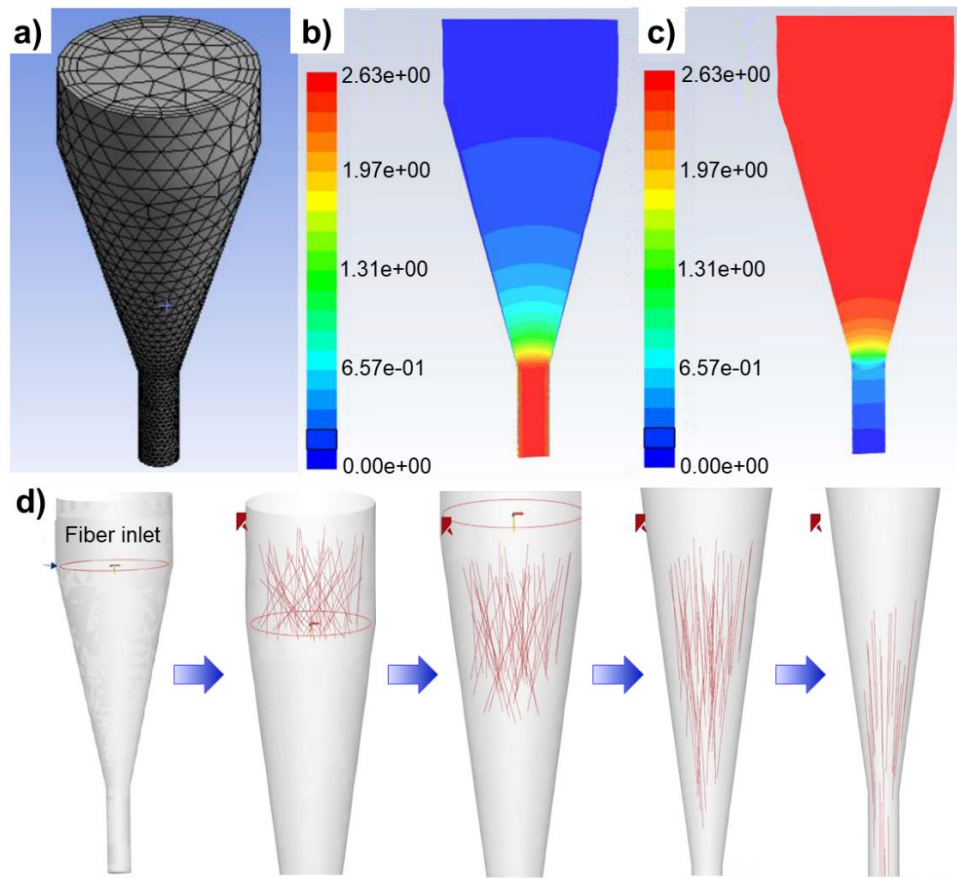

**Figure 4** Theoretical simulations of shear-alignment of CNTs in epoxy matrix. (**a**) The grid used for the calculation of the CNTs-epoxy flow with inlet velocity of 0.1m/s and zero outlet pressure. The calculated velocity (**b**) and pressure (**c**) distribution. An inlet is created as shown in Figure (**d**) which will initiate particles inside the domain. The flow of the CNTs in the epoxy at various stages is shown in figures (**d** ).

To better understand the shear-alignment of CNTs in epoxy during printing we performed theoretical simulation of the flow of CNTs in epoxy using coupled simulations for fluid and fiber interactions. The flow of epoxy through a nozzle is first simulated using ANSYS-Fluent™, and the interaction of CNTs with the flow is further simulated using the Discrete Element Method

(DEM) in the software Rocky-DEM™. One-way interaction of the fluid and fiber is considered where the fluid flow impacts the fiber, but itself remains unaffected by the presence of fiber. **Figure 4a** shows the grid used for the calculation of the epoxy flow with an inlet velocity of 0.1 m/s and zero outlet pressure. The calculated velocity and pressure distribution are shown in **Figure 4b** and **4c**. Velocity values from Fluent results are imported in RockyDEM to calculate fluid forces on CNTs. An inlet is created as shown in **Figure 4(d)** which will initiate particles inside the domain. CNTs are modelled as flexible fibers and default values for material properties are used with the aspect ratio of 200 for the CNTs dimensions. Relatively small number of CNTs have been considered in the simulation for computational tractability. CNTs are generated with a random orientation in the 3D space and their flow in the epoxy at various stages is shown in **Figures 4d**. Initially the CNTs move relatively slowly with the flow of epoxy. As the area of the nozzle starts reducing, the CNTs-epoxy mixture is exposed to a high shear rate and pressure, along with a turbulent environment, enabling close contact between the polymer matrix and CNTs and resulting in alignment of the CNTs in the tapered area of the nozzle before filament extrusion. As can be seen in **Figure 4d,** the CNTs are well aligned with the flow at the end of the nozzle which is consistent with the SEM images discussed previously. The video of the flow of CNT in the epoxy is found in the supplementary information.

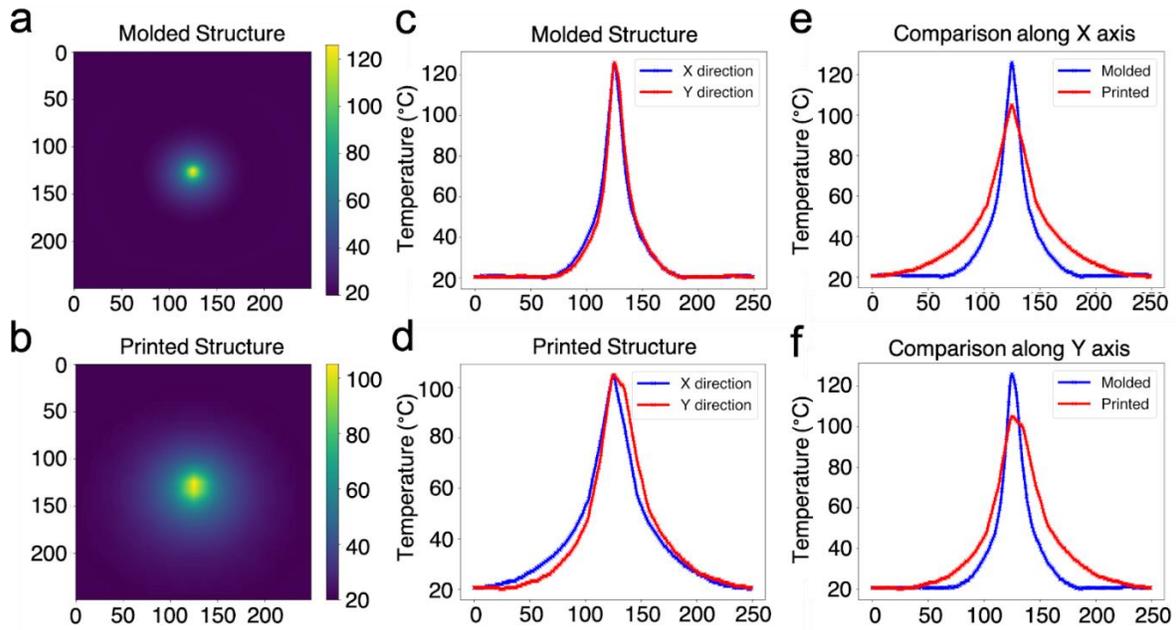

**Figure 5** Thermal map of mold cast **(a)** and 3D-printed **(b)** structure under 45 mW laser illumination. Line profile of the temperature along the X and Y direction of the mold cast **(c)** and 3D-printed **(d)** composites. The line profile of the surface temperatures between the two structures were further compared along X axis **(e)** and Y axis **(f)**.

Next, we compare the thermal behavior (heat transfer) of the mold-casted and printed CNT-epoxy composites to understand the effect of above-discussed structural variations. CNTs are often used for their thermal conductivity and ability to dissipate heat quickly, however, it is known that voids in composite materials and isotropic ordering throttles the full potential of CNTs composites.[33] To investigate the thermal conductivity, the *in-situ* surface temperatures of both 3D printed and mold casted samples were recorded with a thermal IR camera, while white light from a supercontinuum laser is applied for light-induced heating. For comparison, the single line temperature profile along different directions on the sample surface were extracted from temperature mapping captured by the IR camera for both samples. The results of the thermal camera measurements are summarized in **Figure 5**. For both samples, no obvious difference can be seen between the temperature profile along X and Y directions. This indicates that the heat conducting behavior of both samples is homogenous along different directions. Therefore, the alignment of CNTs in the printed structures has little influence on the heat conducting behavior. However, when illuminated with the laser of same power, the line profile of the molded structure showed higher peak temperature but smaller full width at half maximum (FWHM) than that of the printed structure. This shows that the printed samples have superior thermal conductivity and heat dissipation properties compared to the mold-cast samples likely due to their lack of voids and anisotropic CNTs alignment.

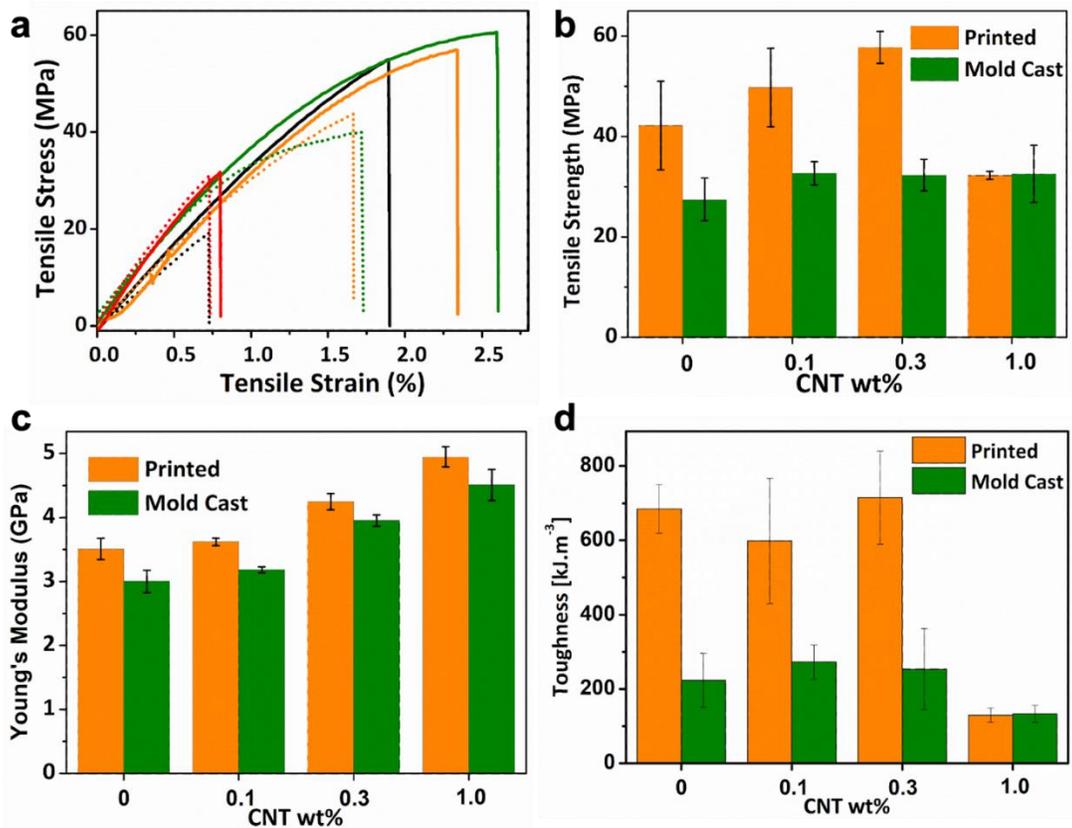

**Figure 6 (a)** Typical tensile stress vs strain plots for the different CNTs-epoxy samples. Solid lines indicate printed sample and dotted lines indicate mold cast counterparts. Colors for different CNT wt.%: black=0, orange=0.1, green=0.3, red=1.0. **(b)** Tensile strength of the different printed and mold cast samples with error bars. **(c)** Average Young's modulus of the different printed and mold cast samples with error bars. **(d)** Calculated average toughness of the different printed and mold cast samples.

Next, mechanical performance of the mold-cast and 3D printed samples are analyzed and compared (**Figure 6a**). The 3D printed composite exhibits higher tensile strength and fracture strain with increasing CNTs loading compared to its mold cast counterpart. This enhancement is attributed to the improved processing dynamics of CNTs-epoxy composite such as- the reduction in voids, homogeneous distribution, better alignment, wettability, and adhesion of the CNTs with the epoxy matrix, thanks to the direct ink writing technique that induces shear forces on the ink during extrusion through the micronozzle. This is further corroborated by the observed differences in fracture surface roughness of the printed and mold casted sample shown in Figure 2c and 2d. The relatively smooth surface of the mold casted sample indicates generally uninterrupted crack propagation after a crack has been initiated. Therefore, it can be said that crack propagation in these samples occurs with less interruption. On the other hand, for 3D printed samples, the fracture surface appeared to have a greater degree of roughness, which can be attributed to more crack deflection and propagation in different fracture planes due to the uniform presence of CNTs. Note that, increased CNTs loading improves the strength of samples regardless of fabrication method (**Figure 6c**). Additionally, as expected, the 3D printed samples have higher Young's modulus than their respective mold cast counterparts. Similarly, the tensile strength and toughness is higher in 3D printed samples and increases with CNT loading, except in the case of 1 wt.% CNT loading, where a rapid drop off in mechanical properties is observed for both fabrication methods (**Figure 6b,d**). We attribute this to incomplete photocuring due to the high optical absorption of CNT (>95 %) preventing UV light from fully penetrating the sample. Overall, improved mechanical properties are observed in the 3D printed structures likely due to partial anisotropic alignment of the CNTs, better epoxy/CNT bonding, and a reduction of voids in the epoxy.

## 3. Conclusions

In summary, complex architected and mechanically, and thermally robust CNTs-epoxy composites were 3D printed from engineered viscoelastic inks consisting of CNTs, epoxy, and rheology modifier. This work demonstrates the potential of direct ink writing technology to mitigate the critical processing factors such as poor interfacial adhesion, wettability, and agglomeration of CNTs-polymer (epoxy, in this case) composite fabrication. We show that direct ink writing technique provide a better microstructure by tackling such critical factors compared to the traditional mold cast technique. Consequently, compared to their mold cast counterparts, the printed samples show significantly higher Young's modulus, fracture strengths, and toughness, while exhibiting superior heat transfer mechanisms. This work demonstrates that direct ink writing is not only useful in fabricating complex on-demand structures, but also offers advantages for composite fabrication by intrinsically improving the micro-structure and physical properties. This capability adds a new dimension to engineering design and process optimization for realizing high-performance CNT-polymer composites.

**Supporting Information**

Supporting Information is available from the Wiley Online Library or from the author.

**Acknowledgements**

The authors gratefully acknowledge financial support from Aramco Research Center (Grant Reference Number: 1137681).

**Conflict of Interest**

The authors declare no conflict of interest.

**Keywords**